\documentclass[twocolumn,showpacs,preprintnumbers,amsmath,amssymb]{revtex4}

\usepackage[pdftex]{graphicx}
\usepackage{dcolumn}
\usepackage{bm}

\begin{document}


\title{Size dependence of the bulk and surface phonon modes of gallium arsenide nanowires as measured by Raman Spectroscopy}

\author{Dan\v{c}e Spirkoska, Gerhard Abstreiter and Anna Fontcuberta i Morral}

\affiliation{Technische Universit\"at M\"unchen, Walter Schottky Institut, Am Coulombwall 3, 85748 Garching, Germany}

\date{\today}

\begin{abstract}
Gallium arsenide nanowires were synthesized by gallium-assisted molecular beam epitaxy. By varying the growth time, nanowires with diameters ranging from 30 to 160 nm were obtained. Raman spectra of the nanowires ensembles were measured. The small line width of the optical phonon modes agree with an excellent crystalline quality. A surface phonon mode was also revealed, as a shoulder at lower frequencies of the longitudinal optical mode.  In agreement with the theory, the surface mode shifts to lower wave numbers when the diameter of the nanowires is decreased or the environment dielectric constant increased.
\end{abstract}

\maketitle

\begin{figure}[hbtp]
\centering
\includegraphics{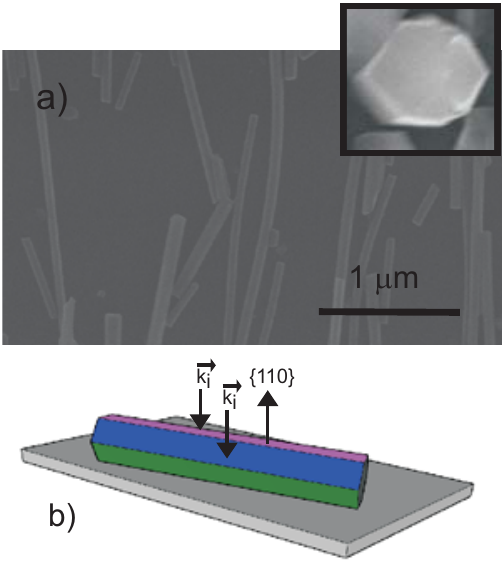}
\caption{a) SEM picture from GaAs nanowires transferred on Si substrate. After the transfer the wires are partially oriented. The hexagonal cross section of the nanowires is presented in the inset. b) Schematic drawing of the scattering geometry of the measurements.}
\label{fig:Figure1}
\end{figure}

\begin{figure}[hbtp]
\centering
\includegraphics{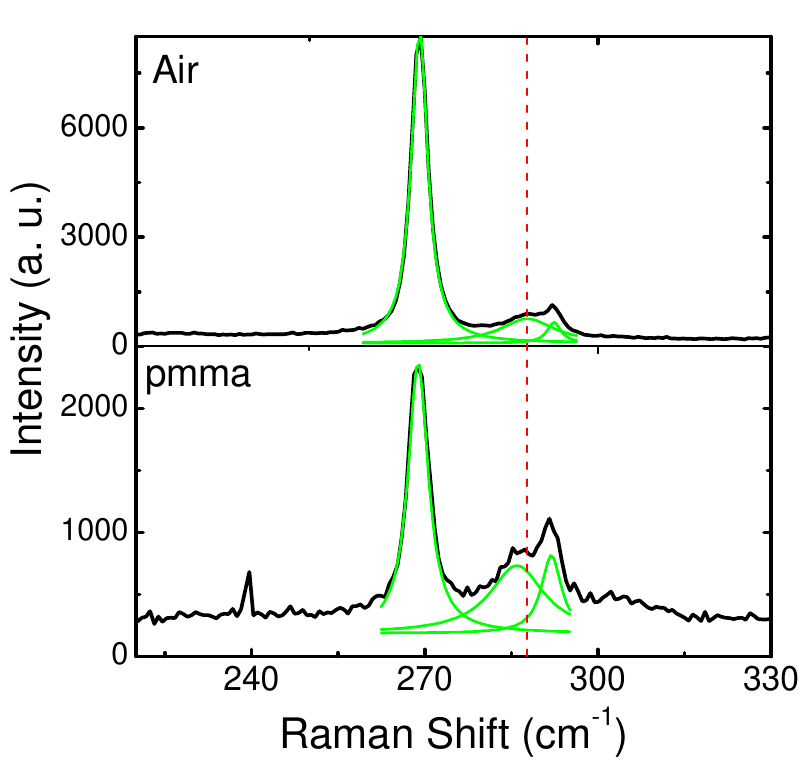}
\caption{Raman spectra of GaAs nanowires with average diameter of 89 nm. The black solid line is the recorded data while the green lines are result from multiple Lorentzian fit. The bottom spectrum corresponds to the measurement realized on the same GaAs nanowires after embedding them in pmma. The mode attributed to surface phonons presents clearly a shift.}
\label{fig:Figure2}
\end{figure}

In the past decade the field of semiconducting nanowires has developed significantly mostly due to the fact that these systems with unique geometry offer great possibilities for further development of optic and electronic devices$^{1,2}$. Equally important, they offer numerous possibilities for studying exciting physical phenomena arising from carrier confinement and/or the large surface-to-volume ration$^{3}$. However, the growth of nanowires free of structural defects and contaminants is still one of the key issues.\\ 
The vapor-liquid-solid growth method is one of the most common technique, in which typically gold is used as a catalyst for the nucleation and growth of the nanowires$^{4}$. It is widely known that gold introduces deep level traps in the semiconductor band gap that hinder the opto-electronic properties of the material$^{5}$. By avoiding the use of gold, the properties of the grown nanowires improve significantly. Recently, catalyst-free synthesis of III-V nanowires has been demonstrated by both MOCVD and MBE$^{6,7}$. From the two techniques, with MBE it is possible to obtain materials with an extremely high purity and good structural properties.
\\Raman spectroscopy as a non destructive characterization tool is extensively applied for characterization of low dimensional systems such as nanowires and nanocrystals, as it provides valuable information on the structural properties$^{8,9}$.\\
In this letter we present a systematic Raman spectroscopy study of GaAs nanowires grown by MBE without the use of external catalyst for the growth. Nanowires with diameters ranging from 30 nm up to 160 nm were grown. The underlaying motivation for investigation of nanowires with a broad range of diameters was to correlate the features in the Raman spectrum with the change of the surface to volume ratio.\\
The nanowires were grown in a GEN II MBE system. For the growth we have used (001) GaAs wafers covered with a thin layer (approximately 35 nm) of sputtered $SiO_{2}$. In order to ensure clean surface, prior to growth the $SiO_{2}$ thin films were etched down to 10 nm by a diluted buffered HF solution. After the etching the substrates were blown dry with nitrogen and immediately transferred in the MBE system. Prior to growth, the substrates were heated to a temperature of 650 $^{\circ}$C in order to desorb any remnant molecules on the surface. Then, the temperature was lowered to the growth temperature of 630 $^{\circ}$C. For the growth we have used As beam equivalent pressure (BEP) of 2$\times$10$^{-6}$ mbar and Ga growth rate of 0.25 \r{A}/s, which gives respectively longitudinal and radial growth rates of 2\r{A}/s and 0.07 \r{A}/s$^{10}$. The nominal thickness of deposited GaAs was varied for different samples in order to synthesize nanowires with different average diameter. In this way, we have prepared samples with diameters ranging from 30 nm up to 160 nm. Each sample was characterized with rather narrow diameter distribution below 10\%.\\
Transmission electron microscopy (TEM) analysis on the grown wires showed that the wires grow in the $(1\overline{1}1)$ B growth direction and have a hexagonal cross section with side facets belonging to the $\left\{110\right\}$ crystalline family$^{11}$. AFM measurements on single nanowires have also shown that the side facets exhibit very small roughness.\\ The Raman measurements were performed at room temperature by using the 488 nm line from Ar$^{+}$ laser. A microscope objective (50x) focused the laser on the sample with a spot of several micrometers in size. The same lens collected the scattered light to a triple DILORXY spectrometer and was further analyzed with a nitrogen cooled Si CCD. The measurements were realized with low excitation power (0.5 mW), with the purpose of avoiding the heating of the sample, which can produce asymmetric broadening and down shift of the Raman peaks. The sample temperature stayed always below 120 $^{\circ}$C, as shown by Stokes/Antistokes ratio measurements.\\
The wires were mechanically removed from the GaAs substrate and transferred on clean (001) Si pieces by friction between the substrates. The transferred wires were partially oriented along the sliding direction, as shown in the scanning electron micrograph presented in Figure \ref{fig:Figure1} a). The inset shows the hexagonal cross section of the wires. Since the laser was focused on a spot with diameter of several microns, we estimate that only several nanowires were probed during each measurement. The scattering geometry is presented on Figure \ref{fig:Figure1} b). The wires are probed in back scattering geometry with the side facets belonging to the $\left\{110\right\}$ crystalline family. In the backscattering geometry from $\left\{110\right\}$ the longitudinal optical (LO) phonon is forbidden according to the Raman selection rules$^{12}$. In principle, only scattering from the transverse optical phonon (TO) is allowed. As a consequence, the scattering from the top and bottom surfaces of the wires will contribute only in the TO signal in the Raman spectra. On the other hand the scattering due to LO phonon is not forbidden from the facets painted in blue. The scattering in these facets will contribute to the LO signal in the spectra. As a result, the spectra should present rather high component of the TO signal and a lower LO signal.\\
A typical Raman spectra of the nanowires (with average diameter of 89 nm) is presented in the upper graph in Figure \ref{fig:Figure2}. The solid black line is the recorded data while the green lines are result from a multiple Lorentzian fit. 
On the spectra from the wires two peaks can be clearly observed. The peak positioned at 268.7 $cm^{-1}$ is due to scattering from TO phonon and the peak positioned at 292.2 $cm^{-1}$ is due to scattering from LO phonon. The TO and the LO peaks are symmetric and have very small FWHM (around 4 $cm^{-1}$). The peak position corresponds exactly with the position of the TO and the LO peaks measured on bulk (111) GaAs$^{13}$ . The measured values for the peak positions and FWHM indicate that the synthesized wires have excellent structural quality and are free of defects and stress, which further corroborates the advantage of using MBE. 
\\A third peak positioned at the low frequency side from the LO phonon is also clearly observed. A detailed analysis of this peak, which will be presented in the following allowed us to attribute it to scattering from surface optical phonon (SO). A simple mathematical expression of the surface modes is given in the case of an infinitely long cylinder$^{14,15}$:
\begin{equation}
\omega^{2}=\omega^{2}_{TO}\frac{(\epsilon_{0}-\epsilon_{m}\rho)}{\epsilon_{\infty}-\epsilon_{m}\rho}
\label{eq:ravenka1}
\end{equation}
\\where 
\begin{equation}
\rho=(K^{'}(hr)I(hr))/(K(hr)I^{'}(hr))
\end{equation}
\\with I(hr) and K(hr) being the modified Bessel functions, h is the propagation constant trough the cylinder, $\epsilon_{0}$ and $\epsilon_{\infty}$ are the static and the high frequency dielectric constant of the material while $\epsilon_{m}$ is the dielectric constant of the surrounding medium.
\\The position of the SO depends from the dielectric constant of the medium that is surrounding the wires as well as from the diameter of the wires. In order to confirm the nature of the third mode as a surface phonon, we have embedded the nanowires with pmma (polymethyl methacrylate), which has a dielectric constant of 2.8. As it can be seen in the bottom panel of Figure \ref{fig:Figure2} there is indeed a significant shift of 1.25 $cm^{-1}$ of the SO phonon toward lower wave numbers. 
\begin{figure}[hbt]
\centering
\includegraphics{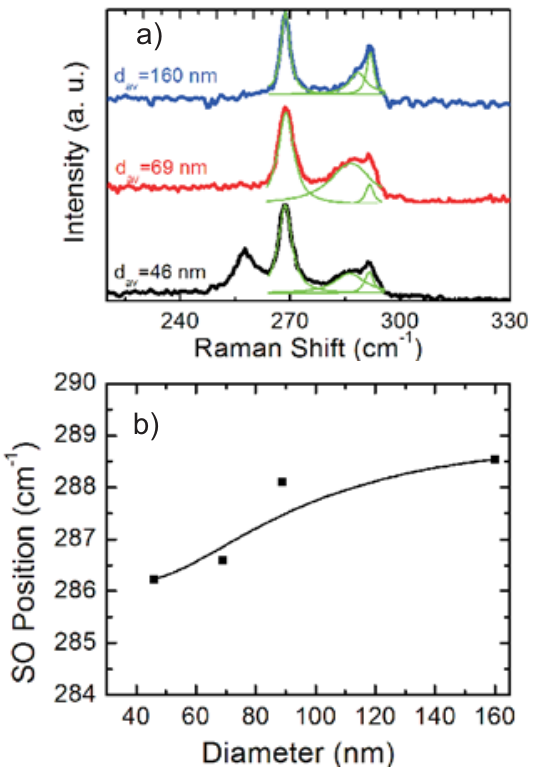}
\caption{a)Raman spectra from GaAs nanowires with average diameters from 46 nm (black), 69 nm (red) and 120 nm (blue). The green lines are result from multiple Lorentzian fit. The position of the SO phonon down shifts with the decrease in the diameter. b) Dependence of the position of the SO phonon from the diameter of the nanowires. The points represent experimental data obtained from several measured samples with different average diameter. The line is a guide to the eye.} 
\label{fig:NewFigure3}
\end{figure} 
\\We have measured the Raman spectra of the other samples consisting of nanowires with diameters from 30 to 160 nm. Previous experimental reports on surface phonons in III-V nanowires and nanocolumns showed a down shift of the position of the SO phonon with a decrease in the diameter of the nanowires$^{14,16}$. On Figure \ref{fig:NewFigure3} a) we present representative measurements performed on samples with average diameters of 160, 69 and 46 nm.
\\As shown the figure, the position of the TO and the LO phonons is independent on the nanowire diameter. However, the SO modes exhibit a down shift when the diameter decreases. Additionally, for thick nanowires with an average diameter of 160 nm, the SO appears only as a small shoulder on the lower frequency side of the LO peak. However, for nanowires with smaller diameters, meaning a larger surface-to-volume ratio, the SO becomes more pronounced and down shifts. This trend further proves that the peak located between the TO and the LO peaks can indeed be attributed to scattering from surface phonons.\\In the Raman spectra from the wires with the smallest diameter (black curve) another peak appears at around 258 $cm^{-1}$. The appearance of these peak in Raman spectra of GaAs has been assigned to the presence of As anti site defects$^{17}$. However, we believe that  this peak could also be attributed to the existence of twins. Indeed, high densities of twins are observed in the nucleation and final stage of growth. As the small diameter wires have a length of 500 to 800 nm, the twinned region becomes up to 20 \% of the wire. More detailed investigations on the origin of the peak are currently ongoing.
\\In order to get a clear overview on \ref{fig:NewFigure3} b) we have plotted the dependence of the position of the SO phonon from the diameter of the nanowires for several measured samples with different average diameter. By comparing our experimental data points with the theoretical curve calculated in ref. 12 according to Equation 1 one can see that we observe a bit larger shift in the SO phonon position. One possible reason can be the fact that our wires have hexagonal cross section. Lately it was shwon that the cross sectional shape of the nanowires is important when calculating the dispersion curve for the SO phonon$^{18}$ . Of course this opens a route for making theoretical calculation for the SO dispersion curve for hexagonal geometry.
\\In summary we have presented Raman spectroscopy study of GaAs nanowires synthesized in MBE by the Ga assisted growth method. We have presented that the nanowires exhibit very good structural quality shown by the line shape and position of the peaks. Furthermore we have demonstrated the existence of a surface phonon mode by comparing the spectra of the nanowires in environment of air and pmma, which has a higher dielectric constant. By investigating spectra of nanowires as a function of the average diameters, we have found that the position of the SO phonon down shifts with the decrease of the diameter.
\\\textbf{Acknowledgments:} The authors kindly thank M. Stuzmann, R. Gross, J. R. Morante and B. Laumer for experimental support and discussions, as well as the funding from Marie Curie Excellence Grant SENFED, Nanosystems Initiative Munich (NIM) and SFB 631.\\
\textbf{References:}\\
\small{
\\$^{1}$ S. De Franceschi, J. A. van Dam, E. P. A. M. Bakkers, L. F. Feiner, L. Gurevich and L. P. Kouwenhoven, Appl. Phys. Lett. \textbf{83}, 344 (2003).\\
$^{2}$ V. Schmidt, H. Riel, S. Senz, S. Karg, W. Riess and U. Goesle, Small \textbf{2},85(2006).\\
$^{3}$ W. Lu, J. Xiang, B. P. Timko, Y. Wu and C. M. Lieber, PNAS \textbf{102}, 10046 (2005).\\
$^{4}$ R. S. Wagner and W. C. Ellis, Appl. Phys. Lett. \textbf{4}, 89 (1964).\\
$^{5}$ S. D. Brotherson and J. E. Lowther, Phys. Rev. Lett. \textbf{44}, 606 (1980).\\ 
$^{6}$ B. Mandl, J. Stangl, T. Martensson, A. Mikkelsen, J. Erkisson, L. S. Karlsson, G. Bauer, L. Samuelson, and W. Seifert, Nano Letters \textbf{6}, 1817 (2006).\\
$^{7}$ R. Banerjee, A. Bhattacharya, A. Genc, and B. M. Arora, Philos. Mag.
Lett. \textbf{86}, 807 (2006).\\
$^{8}$ C. Steinebach, R. Krahne, G. Biese, C. Sch\"uller, D. Heitmann and K. Eberl, Phys. Rev. B, \textbf{54}, R14281 (1996).\\
$^{9}$ R. Krahne, G. Chilla, C. Sch\"uller, L. Carbone, S. Kudera, G. Mannarini, L. Manna, D. Heitmann, and R. Cingolani, Nano Letters, \textbf{6}, 478 (2006).\\ 
$^{10}$ C. Colombo, D. Spirkoska, M. Frimmer, G. Abstreiter and A. Fontcuberta i Morral, Phys. Rev. B., in press (2008).\\
$^{11}$ A. Fontcuberta i Morral, D. Spirkoska, M. Heigoldt, J. Arbiol, J. R. Morante and G. Abstreiter, Small, in press (2008).\\
$^{12}$ M. Cardona and G. Guntherodt, Light Scattering in Solids, Springer.\\
$^{13}$ A. Mooradian and G. B. Wright, Solid State Communications, \textbf{4}, 431 (1966).\\
$^{14}$ M. Watt, C. M. Sotomayor Torres, H. E. G. Arnot and S. P. Beaumont, Semicond. Sci. Technol. \textbf{5}, 285 (1990).\\
$^{15}$ R. Ruppin and R. Englman, Rep. Prog. Phys. \textbf{33}, 149 (1970).\\
$^{16}$ R. Gupta, Q. Xiong, G. D. Mahan and P. C. Eklund, Nano Letters, \textbf{3}, 1745 (2003).\\
$^{17}$ M. Touffela, P. Puech, R. Carles, E. Bedel, C. Fontaine, A. Claverie and G. Benassayag, J. of Appl. Phys, \textbf{85}, 2929 (1999).\\
$^{18}$ K. W. Adu, Q. Xiong, H. R. Gutierrez, G. Chen and P C. Eklund, Appl. Phys. A \textbf{85}, 287 (2006).}

\end{document}